\documentclass[twocolumn,aps,prl,showpacs,preprintnumbers,amsmath,amssymb, superscriptaddress]{revtex4-2}
\usepackage{bm}
\usepackage{graphicx}
\usepackage{xcolor,xspace}
\usepackage{ulem}
\usepackage{float}
\usepackage{comment}

\newcommand{\RNum}[1]{\uppercase\expandafter{\romannumeral #1\relax}}

\newcommand{\Rom}[1]{ \uppercase\expandafter{\romannumeral#1}}

\definecolor{ZZYcolor}{rgb}{0.1,0.5,0.4}

\newenvironment{addendum}{%
   \setlength{\parindent}{0in}%
   \small
   \begin{list}{Acknowledgements}{%
       \setlength{\leftmargin}{0in}%
       \setlength{\listparindent}{0in}%
       \setlength{\labelsep}{0em}%
       \setlength{\labelwidth}{0in}%
       \setlength{\itemsep}{12pt}%
       }
   }
   {\end{list}\normalsize}

\begin{document}

\title{Evolution of the intertwining correlated topological phases in iron-based superconductor Fe(Te,Se)}

\author{Yue Sun}\thanks{These authors contributed equally to this work.}
\affiliation{Key Laboratory of Quantum Materials and Devices of Ministry of Education, School of Physics, Southeast University, Nanjing 211189, China}

\author{Shiying He}\thanks{These authors contributed equally to this work.}
\affiliation{National Laboratory of Solid State Microstructures, School of Physics , Nanjing University, Nanjing 210093, China.}

\author{Zhongyi Zhang}\thanks{These authors contributed equally to this work.}
\affiliation{Department of Physics, Hong Kong University of Science and Technology, Clear Water Bay, Hong Kong, China}

\author{Yong Huang}
\affiliation{National Laboratory of Solid State Microstructures, School of Physics , Nanjing University, Nanjing 210093, China.}

\author{Jingheng Chen}
\affiliation{National Laboratory of Solid State Microstructures, School of Physics , Nanjing University, Nanjing 210093, China.}

\author{Weixiang Yan}
\affiliation{National Laboratory of Solid State Microstructures, School of Physics , Nanjing University, Nanjing 210093, China.}

\author{Chunbo Yu}
\affiliation{National Laboratory of Solid State Microstructures, School of Physics , Nanjing University, Nanjing 210093, China.}

\author{Yuyang Dong}
\affiliation{Institute for Solid State Physics, University of Tokyo, Kashiwa, Chiba 277-8581, Japan}

\author{Kohei~Aido}
\affiliation{Institute for Solid State Physics, University of Tokyo, Kashiwa, Chiba 277-8581, Japan}

\author{Xin Zhou}
\affiliation{Key Laboratory of Quantum Materials and Devices of Ministry of Education, School of Physics, Southeast University, Nanjing 211189, China}

\author{Zhengtai Liu}
\affiliation{Shanghai Synchrotron Radiation Facility, Shanghai Advanced Research Institute, Chinese Academy of Sciences, Shanghai 201210, China}

\author{Mao Ye}
\affiliation{Shanghai Synchrotron Radiation Facility, Shanghai Advanced Research Institute, Chinese Academy of Sciences, Shanghai 201210, China}

\author{Jishan Liu}
\affiliation{Shanghai Synchrotron Radiation Facility, Shanghai Advanced Research Institute, Chinese Academy of Sciences, Shanghai 201210, China}

\author{Haruhisa Kitano}
\affiliation{Department of Physics, Aoyama Gakuin University, Sagamihara 252-5258, Japan}

\author{Zhixiang Shi}
\affiliation{Key Laboratory of Quantum Materials and Devices of Ministry of Education, School of Physics, Southeast University, Nanjing 211189, China}

\author{Hong~Ding}
\affiliation{Tsung-Dao Lee Institute and School of Physics and Astronomy, Shanghai Jiao Tong University, Shanghai, 200240 China}
\affiliation{Hefei National Laboratory, Hefei 230088, China}
\affiliation{New Cornerstone Science Laboratory, Shanghai 201210, China}

\author{Takeshi Kondo}
\affiliation{Institute for Solid State Physics, University of Tokyo, Kashiwa, Chiba 277-8581, Japan}
\affiliation{Trans-scale Quantum Science Institute, The University of Tokyo, Tokyo 113-0033, Japan}

\author{Xianxin Wu}
\affiliation{CAS Key Laboratory of Theoretical Physics, Institute of Theoretical Physics, Chinese Academy of Sciences, Beijing 100190, China}
\email{xxwu@itp.ac.cn}

\author{Peng Zhang}
\affiliation{National Laboratory of Solid State Microstructures, School of Physics , Nanjing University, Nanjing 210093, China.}
\affiliation{Collaborative Innovation Center for Advanced Microstructures, Nanjing, China}
\affiliation{Jiangsu Physical Science Research Center, Nanjing, China}
\email{zhangpeng@nju.edu.cn}

\date{\today}

\begin{abstract}

Multiple topological electronic phases can coexist within a single quantum material and induce different topological superconducting states, offering deeper insights into interplay of topological superconducting states and Majorana modes, which may also be influenced and modified by correlation effect.
Iron-based superconductors, with both topological states and correlation effect, is an ideal platform to study these phenomena. 
Here, with high resolution angle resolved photoelectron spectroscopy, we directly resolve two distinct intertwining topological states in iron-based superconductor Co-doped Fe(Te,Se), and study their evolution with electron doping. We identify a region where both topological insulator surface states and topological Dirac semimetal states intersect the Fermi level. 
The topological states are affected by the strong correlation effect and are isolated from trivial bulk states. The evolution between distinct topological phases offers a good opportunity to study various Majorana modes from different superconducting phases according to theoretical analysis. Our findings establish an ideal platform for exploring the interaction between multiple topological superconducting states and the related Majorana modes.

\end{abstract}

\maketitle

In theory, topological superconductors can be realized by placing a topological insulator (TI) in proximity to an s-wave superconductor~\cite{FuPRL2008}. Extensive researches have focused on heterostructures combining conventional s-wave superconductors with topological insulators or semiconductors~\cite{KouwenhovenScience2012,JiaPRL2016,nadjperge2014observation,albrecht2016exponential}. Alternatively, topological superconductivity has been observed on the surface of iron-based superconductor FeTe$_{0.55}$Se$_{0.45}$ using high-resolution angle-resolved photoemission spectroscopy (ARPES), with a superconducting transition temperature $T_{\rm{c}}$ $\sim$ 14.5 K \cite{ZhangScience2018,KanigelPRB2020,JohnsonPNAS2021,LiNM2021,ZXShenPRX2024}. These findings combine the topology and high-$T_{\rm{c}}$ superconductors, establishing a high-$T_{\rm{c}}$ platform for investigating Majorana states.  Zero-bias peaks inside vortex cores are observed in multiple scanning tunneling microscopy (STM) experiments, strongly supporting the presence of Majorana modes \cite{DingScience2018,HanaguriNM2019,DingScience2020,KongNP2019}. Similar Majorana zero modes have also been detected in iron-based superconductors LiFeAs, (Li$_{0.84}$Fe$_{0.16}$)OHFeSe and CaKFe$_4$As$_4$ \cite{GaoNature2022, FengPRX2018, DingNC2020}. Additionally, helical Majorana modes have been proposed to be localized at the hinges between the top and side faces of Fe(Te,Se) \cite{DasSarmaPRL2019}, and a dispersing 1D Majorana channel has been reported at domain walls \cite{MadhavanScience2020}.

In addition to the topologically superconducting states at the surface, first-principles calculations have also predicted the presence of a topological Dirac semimetal (TDS) bulk band above the TI surface Dirac band in iron-based superconductors \cite{WangPRB2015,WuPRB2016}. 
In Co-doped LiFeAs, an impure bulk Dirac semimetal band has been observed, which coexists with topologically trivial bulk bands, and the topological nature of the Dirac semimetal band is overshadowed by the trivial bands \cite{ZhangNP2019}. In contrast, in this work we find the Dirac semimetal band in Fe(Te,Se) is correlated and exceptionally clean, free from surrounding trivial bulk states. 
With electron doping, the surface Dirac band and the bulk Dirac semimetal band could cross the Fermi level at the same time. Such intersection enables the existence of multiple Majorana states from both the surface Dirac band and the bulk Dirac semimetal band in Fe(Te,Se). 
With higher electron doping, when the Dirac point of semimetal band is tuned to the vicinity of Fermi level, the Dirac semimetal band is dominant and two kinds of Majorana modes, one at the 1D hinge and one in vortex cores, could exist.
Thus, depending on doping level, there are more than one scenario to produce multiple Majorana modes in Fe(Te,Se). In either case, the interactions between different Majorana modes could provide more insights on the manipulation of Majorana modes than that of a single Majorana mode~\cite{KouwenhovenScience2012,DingScience2018,GaoNature2022, JiaNature2024}.

\begin{figure}\center
	\includegraphics[width=0.9\linewidth]{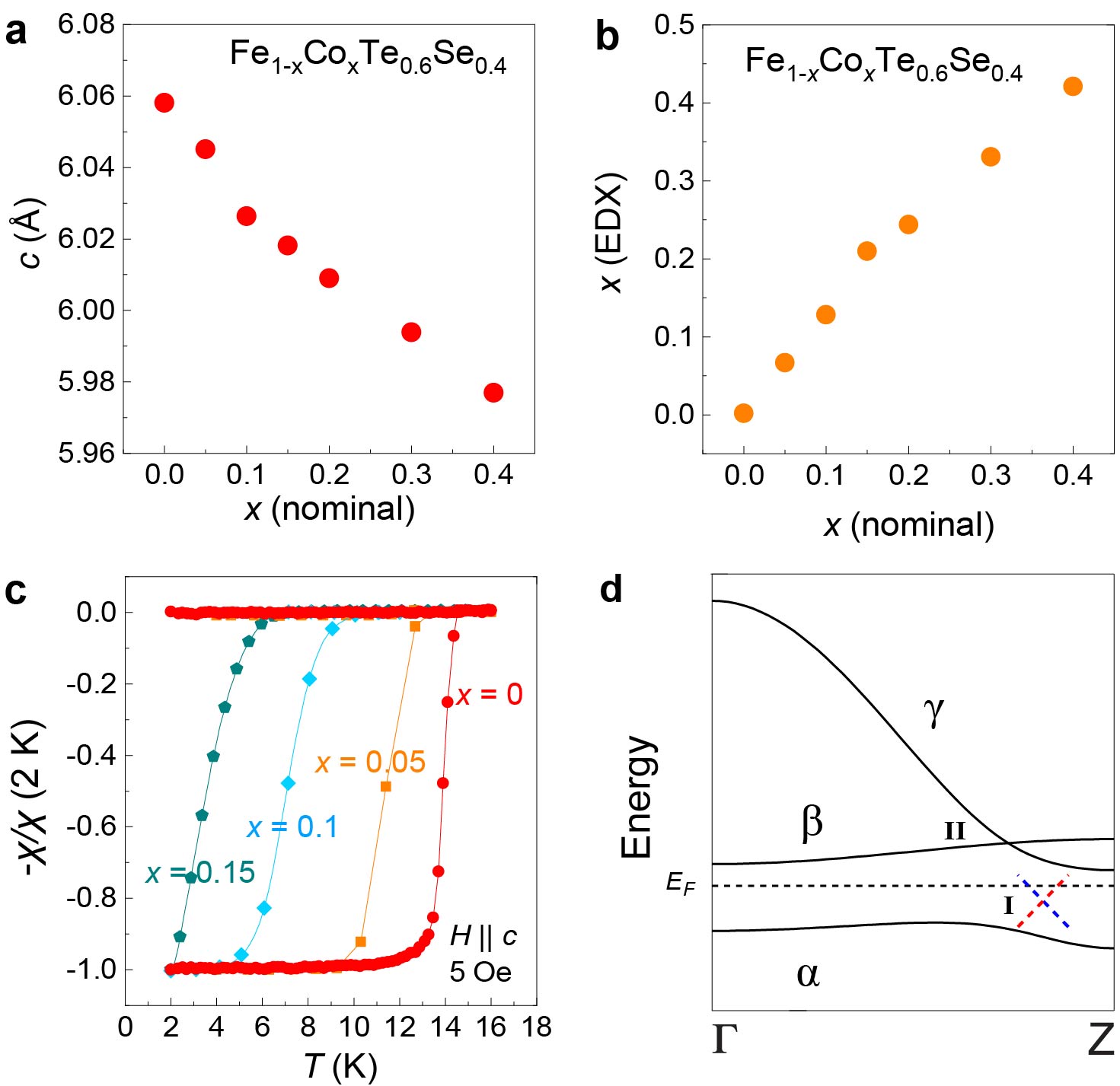}
	\caption{\textbf{Structure, composition and $T_{\rm{c}}$ of Co-doped FeTe$_{0.6}$Se$_{0.4}$. }\textbf{a}, Lattice constant $c$ for Fe$_{1-x}$Co$_x$Te$_{0.6}$Se$_{0.4}$ (0 $\leq$ $x$ $\leq$ 0.4) obtained from the X-ray diffraction measurements. \textbf{b}, EDX results for the actual Co concentrations. \textbf{c}, Temperature dependence of the reduced magnetic susceptibilities -$\chi$/$\chi$(2 K) under 5 Oe for Fe$_{1-x}$Co$_x$Te$_{0.6}$Se$_{0.4}$ (0 $\leq$ $x$ $\leq$ 0.15). \textbf{d}, Sketch of band dispersion along $k_z$ direction. There are two band inversions (Band inversion I and II). $\alpha$/$\beta$ band consists of $d_{xz/yz}$ orbital, while $\gamma$ band mainly consists of $p_z$ orbital. In undoped Fe(Te,Se), Fermi level is located roughly at the dashed line, in the SOC gap of the band inversion I. }\label{}
\end{figure}

\begin{figure*}\center
	\includegraphics[width=0.85\linewidth]{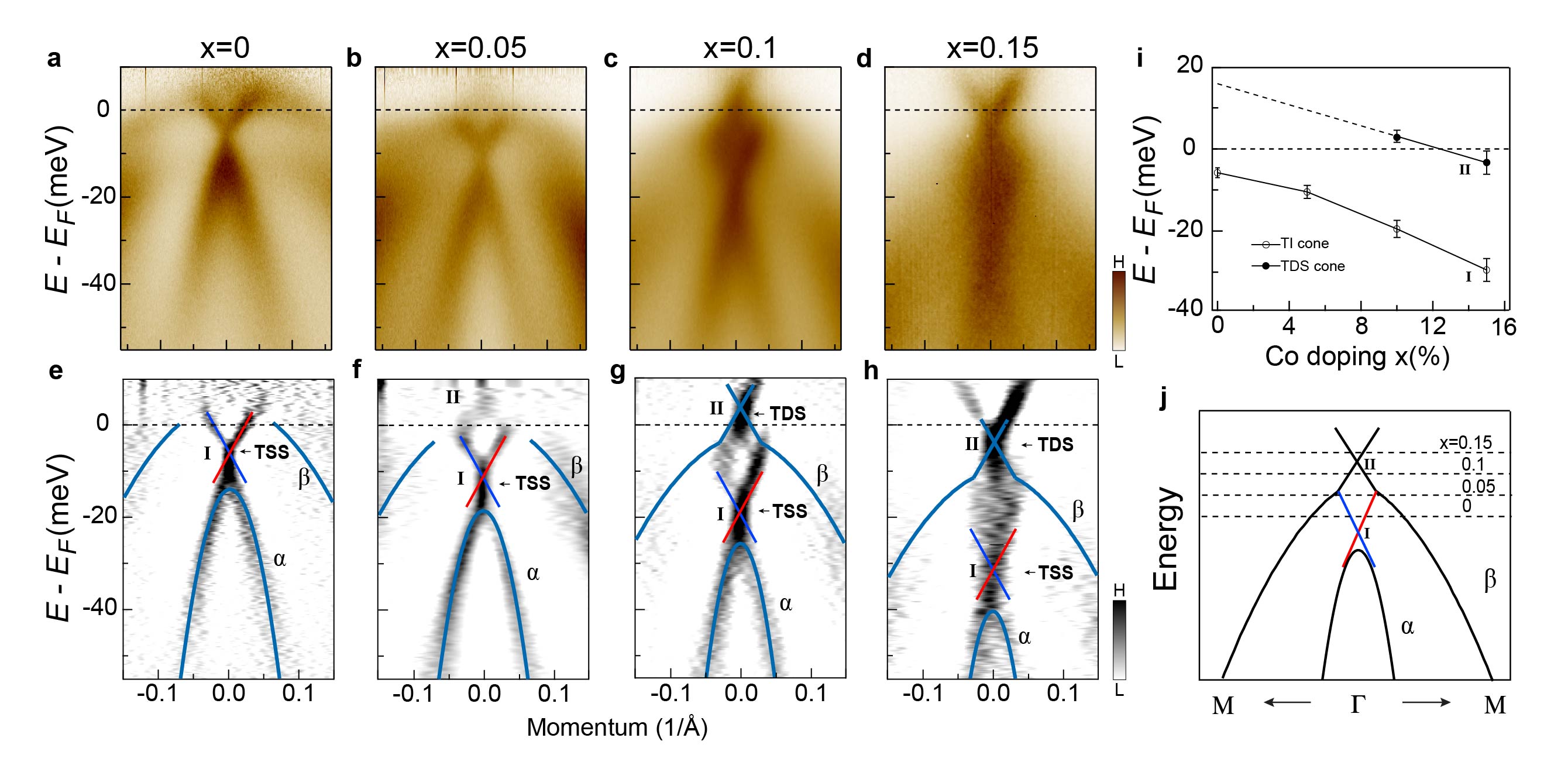}
	\caption{\textbf{Evolution of the TI and TDS Dirac bands with Co doping. }\textbf{a-d}, Band structure of Fe$_{1-x}$Co$_x$Te$_{0.6}$Se$_{0.4}$ ($x$ = 0, 0.05, 0.1, 0.15) near $\Gamma$ at 30 K, with a laser delivering $p$-polarized 7-eV photons. \textbf{e-h}, Curvature intensity plots of \textbf{a-d}. \textbf{i}, Shift of the TI (I) and TDS (II)  Dirac points with Co doping. Extraction method of the Dirac point positions can be found in  Supplementary Information, Section S2. \textbf{j}, Sketch of the in-plane band structure along $\Gamma$M. The Dirac band (I) with red and blue lines is the TI Dirac band. The Dirac band (II) on top is the TDS band. The dashed lines represent the corresponding Fermi level at different Co doping. }
\end{figure*}

Single crystals with nominal compositions Fe$_{1-x}$Co$_x$Te$_{0.6}$Se$_{0.4}$ (0 $\leq$ $x$ $\leq$ 0.4) were grown with flux method and annealed in the atmosphere of Te. More details can be found in Ref. \cite{HouPNAS, SunJPSJTeannealing}. Laser-based ARPES measurements were performed at the Institute for Solid State Physics, the University of Tokyo, with a laser delivering 6.994-eV photons, and a ScientaOmicron R4000 analyzer. The angle resolution was 0.3\textdegree~and the overall energy resolution was set to $\sim$ 2 meV in ARPES measurements. Synchrotron-based ARPES measurements were performed at the BL03U beamline of the Shanghai Synchrotron Radiation Facility using a Scienta Omicron DA30L electron analyzer, and the overall energy resolution was set to $\sim$ 5 meV.

In practice, electron doping in Fe(Te,Se) is challenging due to excess Fe at interstitial positions \cite{BezusyyPhysRevB.91.100502,rosmus2020effect}. In this study, we successfully achieved electron-doped Fe(Te,Se) by substituting Co at the Fe site, effectively mitigating the influence of excess Fe and at the same time maintaining the high crystal quality.
With increased Co content, the positions of (00l) peaks gradually shift to larger 2$\theta$ (see Supplementary Information, Section S1 \cite{supplement}), The lattice constant $c$ is calculated and plotted in Fig. 1a, which is nearly linearly decreasing with $x$. The measured Co-doping level $x$ by energy-dispersive X-ray spectroscopy (EDX) is presented in Fig. 1b. The values of $x$ are slightly larger than the nominal compositions, and show a nearly linear relation. The X-ray diffraction data in Fig. 1a and EDX results in Fig. 1b confirm the successful doping of Co. For simplicity, we refer to the nominal composition throughout the following text. Fig. 1c shows the temperature dependence of the normalized magnetic susceptibilities -$\chi$/$\chi$(2 K) under 5 Oe for Fe$_{1-x}$Co$_x$Te$_{0.6}$Se$_{0.4}$. Clearly, superconductivity above 2 K has been obtained in the rang of 0 $\leq$ $x$ $\leq$ 0.15. With Co doping, onset $T_{\rm{c}}$ is gradually suppressed from 14 K in $x$ = 0 to about 6 K in $x$ = 0.15.

Since additional electrons are introduced with Co doping, a topological Dirac semimetal band is supposed to gradually shift to the Fermi level ($E_{\rm{F}}$)~\cite{ZhangNP2019}.
As shown in Fig. 1d, the topological Dirac semimetal band comes from the band inversion between $\beta$ band ($d_{xz/yz}$ orbital) and $\gamma$ band ($p_z$ orbital mixed with $d_{xy}$ orbital). There are two band inversions along $k_z$ direction in total: band inversion I ($\alpha$ band and $\gamma$ band) and band inversion II ($\beta$ band and $\gamma$ band). There is a spin-orbit coupling (SOC) gap when $\alpha$ band and $\gamma$ band (band inversion I) cross, which forms a TI state and leads to the widely studied topological states at the surface \cite{ZhangScience2018,KanigelPRB2020,JohnsonPNAS2021,LiNM2021,ZXShenPRX2024}. On the other hand, the crossing between $\beta$ band and $\gamma$ band (band inversion II) is protected by crystal symmetry, resulting in the topological Dirac semimetal band. It is obvious that the Dirac semimetal band is on top of the TI band.  In undoped Fe(Te,Se) the TDS band is above $E_{\rm{F}}$ and can only be visible by electron doping.

From laser-ARPES measurements, the band structure of Fe$_{1-x}$Co$_x$Te$_{0.6}$Se$_{0.4}$ with Co composition $x$ = 0, 0.05, 0.1 and 0.15 are shown in Fig. 2a-d. Indeed, the band structure shifts down with increased Co content, confirming the electron doping effect of Co. In $x$ = 0 sample (Fig. 2a, e), only a Dirac band from topological insulator states (Dirac band I) is observed below Fermi level. In $x$ = 0.05 sample  (Fig. 2b, f), the lower part of the second Dirac band (Dirac band II) from topological Dirac semimetal states touches the Fermi level, and the upper part of the TI Dirac band (Dirac band I) is also touching Fermi level. The two Dirac bands both cross Fermi level in this composition. By increasing Co content to $x$ = 0.1 and 0.15, the TDS Dirac band is fully visible, of which the Dirac point is about 20 meV above the TI one, as displayed in Fig. 2c, d. These results are more clearly displayed in the curvature plot in Fig. 2g, h. This is the first time to resolve the full profile of the TDS band in Fe(Te,Se). We extracted the positions of the Dirac points from Fig. 2a-d (Details can be found in  Supplementary Information, Section S2) and plotted against Co content $x$ in Fig. 2i. Both Dirac points shift down linearly with Co doping, indicating a linear electron doping with Co. The energy difference between  TI Dirac point and  TDS Dirac point is about 20 meV. We plotted the sketches of the band dispersion in Fig. 2j. The relative positions of chemical potential in different Co samples are plotted in the in-plane band structure in Fig. 2j. In undoped Fe(Te,Se), The TI Dirac point is at about -5 meV, while the TDS Dirac point is at about 15 meV. In $x$ = 0.05 sample, the TI state coexist with TDS at $E_{\rm{F}}$. Instead, in $x$ = 0.1 and 0.15 samples, the TDS Dirac band is dominant at the Fermi level.

The TDS Dirac cone is more clearly revealed in the temperature-dependent measurements of the $x$ = 0.1 sample (see Supplementary Information, Section S3 \cite{supplement}). Crucially, in contrast to Li(Fe,Co)As where topological states are overwhelmed by trivial bands \cite{MiaoNC2015,BorisenkoNP2016,ZhangNP2019}, the TDS band in (Fe,Co)Te$_{0.6}$Se$_{0.4}$ is the only feature crossing $E_{F}$, thus dominating the electronic properties. This establishes a clean and ideal platform for studying the superconducting states of a correlated TDS.

\begin{figure*}\center
	\includegraphics[width=0.9\linewidth]{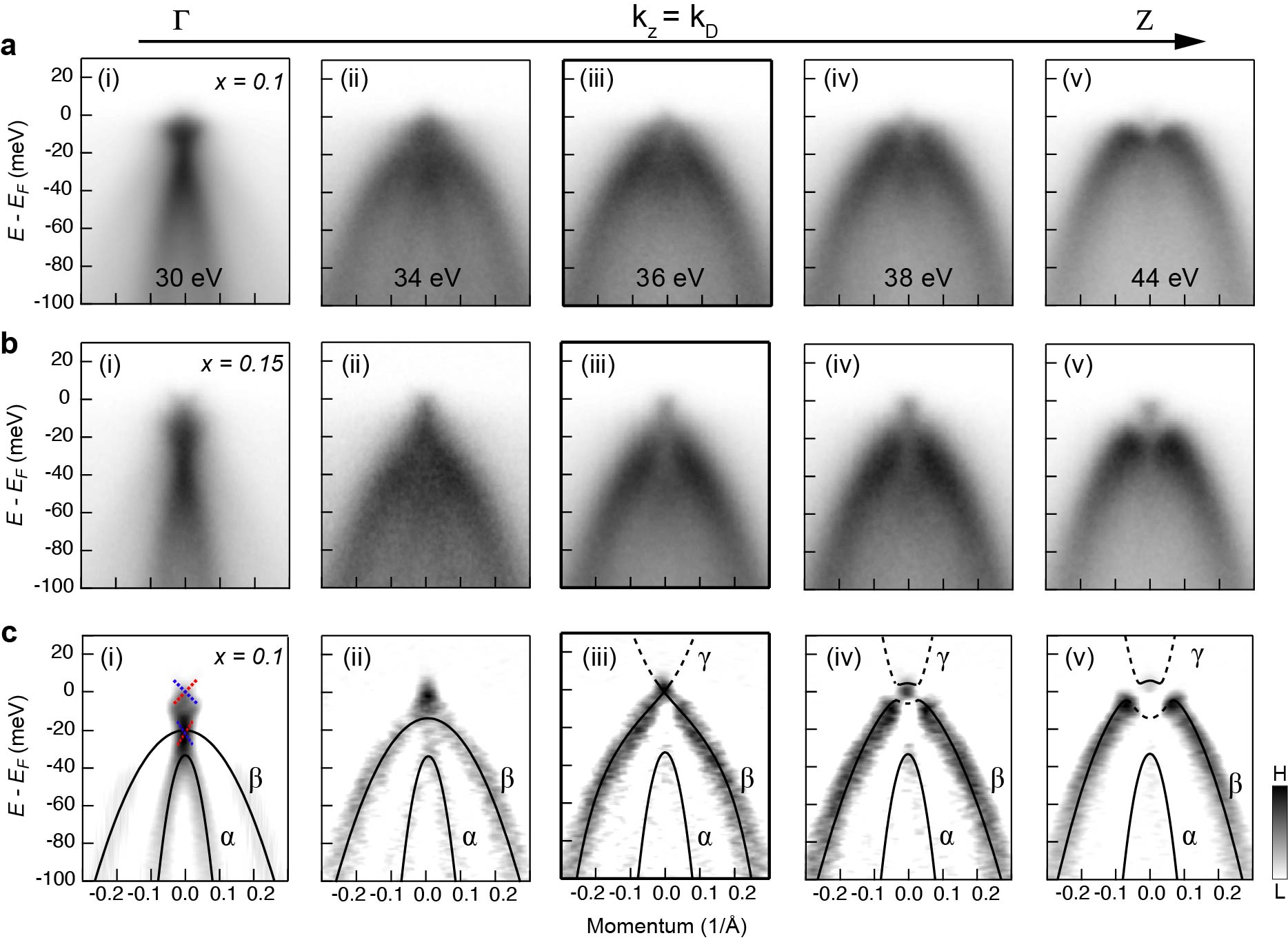}
	\caption{\textbf{Photon-energy dependence of band structure in Fe$_{1-x}$Co$_x$Te$_{0.6}$Se$_{0.4}$ ($x$ = 0.1 and 0.15) at 30K. }\textbf{a}, ARPES spectra of $x$ = 0.1 sample at different photon energies from 30 eV to 44 eV. \textbf{b}, The same as \textbf{a} but with $x$ = 0.15 sample. \textbf{c}, EDC curvature plot of \textbf{a}. The bulk Dirac point is determined to be at around 36 eV. The black lines indicate the bulk bands. The blue and red dashed lines in \textbf{c}(i) indicated the TI surface band and TDS surface components, which are unclear at higher photon energies. }
\end{figure*}

Since the Dirac cone II is from bulk semimetal state, we should be able to resolve its k$_z$ evolution through photon energy dependent measurements. We show the band structures from 30 eV to 44 eV of the $x$ = 0.1 sample and the $x$ = 0.15 sample in Fig. 3a, b. Except the Fermi level shift, the two samples show similar band structure. In Fig. 3c, we display the curvature plots of Fig. 3a and the $\alpha$, $\beta$ and $\gamma$ bands. With different photon energies, $\alpha$ band has almost no shift, consistent with the $k_z$ dispersion illustrated in Fig. 1d, which is due to the SOC hybridization between $\alpha$ band and $\gamma$ band \cite{KanigelPRB2020,ZXShenPRX2024}. In comparison, since there is no hybridization between $\beta$ band and $\gamma$ band, the $k_z$ dispersion of $\beta$ band is preserved (Fig. 1d). Correspondingly, the $\beta$ band in $k_x$-$k_y$ plane should shift upward and $\gamma$ band should shift downward in energy from $\Gamma$ to Z, and they should touch each other at the $k_z$ position of Dirac point. 

Indeed, at the $\Gamma$ point (30 eV), the $\beta$ band is quite below $E_{\rm{F}}$ (Fig. 3c(i)). When increasing photon energy to 44 eV, the $\beta$ band gradually moves up (Fig. 3c(v)). The $\beta$ band shows a parabolic dispersion at 34 eV (Fig. 3c(ii)), which means the two bands are not inverted yet. We observe a almost linear band in the vicinity of Fermi level at 36 eV (Fig. 3c(iii)), which should be the corresponding $k_z$ of the Dirac band II. By further increasing photon energy up to 38 eV and 44 eV, the $\beta$ band has some bending back feature and there is a gap opening between $\beta$ band and $\gamma$ band (Fig. 3c(iv), c(v)), which indicates the two bands are inverted and open a hybridization gap. Thus, we demonstrate the Dirac band II does have a $k_z$ dispersion and confirm its bulk nature. The $k_z$ of the Dirac band II is at about 36 eV. 
We note that at low photon energies ($\le$ 34 eV), both the TI surface band and the surface components of TDS\cite{WangPRB2012,ChenScience2014,JozwiakNC2016,ZhangNP2019} are visible, while they are almost invisible at high binding energies ($>$ 34 eV), which is likely due to the larger cross section of $p$ orbital at low photon energies.

\begin{figure*}\center
	\includegraphics[width=0.65\linewidth]{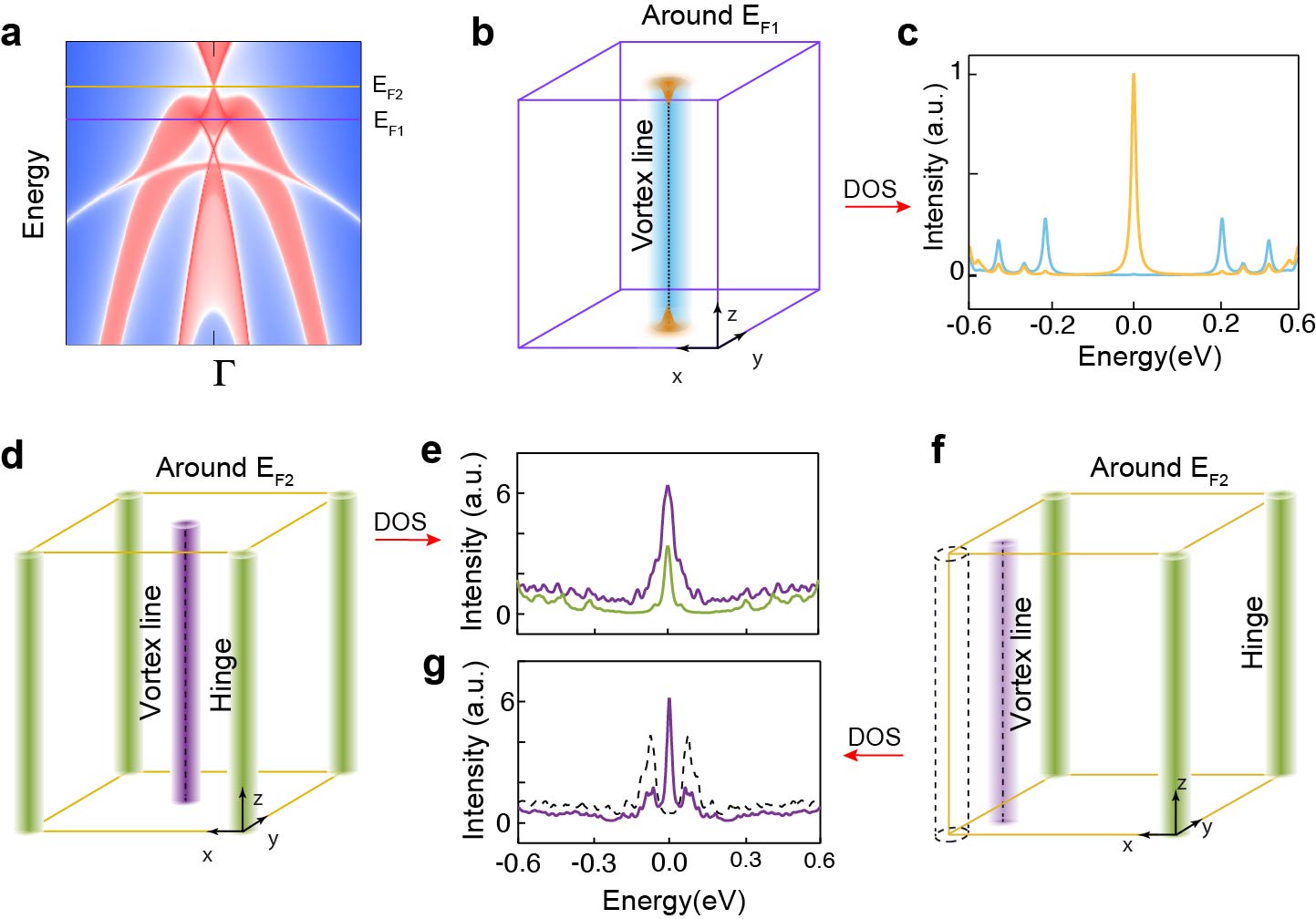}
	\caption{\textbf{Possible Majorana modes in electron-doped Fe(Te,Se)}. \textbf{a}, Schematic surface spectrum of the (001) surface in the normal state.  \textbf{b}, Hybrid vortex phase around $E_\mathrm{F1}$ with conventional $s$-wave pairing, exhibiting both Majorana zero mode (yellow) at the end of vortex and 1D Majorana mode (cyan) along the vortex. \textbf{c}, Calculated local density of states of the two Majorana modes in \textbf{b}. \textbf{d}, \textbf{e}, Majorana mode in the vortex (purple) and Majorana mode at the hinge (green) with $B_{1u/2u}$ pairing around $E_\mathrm{F2}$. \textbf{d} shows the distribution of the two Majorana modes in real space, and \textbf{e} shows their calculated local density of states. \textbf{f}, When the Majorana mode in the vortex is close to a hinge, the two Majorana modes will hybridize, and the one at the hinge will disappear.  \textbf{g}, Calculated local density of states of the vortex and hinge modes in \textbf{f}. The numerical results corresponding to (\textbf{b},\textbf{d} and \textbf{f}) can be found in Supplementary Fig. S8.
 }
\end{figure*}

In LiFeAs, where the correlation effect is relatively weak, the TDS Dirac point (II) is overwhelmed by trivial bulk states~\cite{ZhangNP2019}. In comparison, the correlation level in Fe(Te,Se) is one of the strongest among iron-based superconductors~\cite{PourretPRB2011,KotliarNM2011,ShenPRB2015}. In Supplementary Fig. S6, we show evidences from linear resistivity and large effective mass of $d_{xy}$-orbital band. The strongly correlated $d_{xy}$ orbital hybridizes with $p_z$ orbital, which reduces the $\gamma$ band width~\cite{KimPRL2024}.  In Supplementary Fig. S7, we show this correlation effect drives a dispersion change of the topological bands, producing the clean and isolated TDS Dirac band, which is separated from the trivial bulk states and dominates the electronic density of states when it is located at $E_{\mathrm{F}}$.
The calculated spectrum corresponding to strong correlated $d_{xy}$ orbital and $\gamma$ band is shown in Fig. 4a. The intertwining and clean topological bands lead to a tunable electronic landscape, allowing for the manipulation of topological states: the topological insulating phase evolves to a bulk Dirac semimetal phase under electron doping. 

When the Fermi level is located around $E_{\rm{F1}}$, both TI surface Dirac band and TDS bulk Dirac band are involved. The bulk $d_{yz}$ hole pocket can stabilize $s_{\pm}$-wave pairing. With such a pairing, the TI surface states will contribute Majorana zero modes localized in the vortex, while the bulk Dirac band from TDS will lead to a 1D nodal vortex phase protected by four-fold rotational symmetry \cite{HuPRL2019,ColemanPRL2019}. Together, these elements form a hybrid vortex phase, containing both Majorana bound modes and 1D  Majorana itinerant modes, as illustrated in yellow and cyan colors in Fig. 4b, c, respectively~\cite{ZhangarXiv2025}. When applying a uniaxial strain to reduce the system's symmetry from $C_4$ to $C_2$, the nodal vortex state will be gapped out and change to another Majorana zero mode, yielding two decoupled Majorana zero modes in one vortex. Furthermore, tilting the vortex line by rotating magnetic field will break $C_2$ symmetry, and lead to hybridization of the two different Majorana zero modes. Such hybridization will annihilate both Majorana modes, leaving no Majorana zero mode in vortex.
Further shifting the Fermi level to around  $E_{\rm{F2}}$, the bulk $d_{yz}$ hole pocket vanishes and only Dirac cone from TDS contributes to Fermi surface, realizing an ideal Dirac semimetal phase. The Dirac cone from TDS possesses an unique orbital texture, which dictates the pairing symmetry \cite{SatoPRL2015}: intra-orbital attraction favors conventional intra-orbital pairing, while inter-orbital attraction drives an unconventional orbital-singlet $B_{1u/2u}$ pairing. The latter is particularly relevant in correlated Fe(Te,Se), due to the strong electronic interactions. This $B_{1u/2u}$ pairing features Majorana states at the hinges and Majorana flat bands in vortex, as shown in Fig. 4d, e, which all have sharp zero-biased peaks in the local density of states. When a vortex is very close to a hinge,  the Majorana zero mode in the vortex and that in the hinge will hybridize. As a result, the Majorana hinge mode will disappear, as shown in Fig. 4f, g. 
Therefore, at either $E_{\rm{F1}}$ or $E_{\rm{F2}}$ there is coexistence of different Majorana modes and it is possible to study the evolution and interaction between different Majorana modes by experiments, such as scanning tunneling microscopy. Compared to the research on a single Majorana mode, the characteristics of multiple Majorana modes would provide versatile insights into the braiding operation. 

In summary, we successfully synthesized electron doped Fe(Te,Se), and observed clean topological states from TI and TDS modified by strong correlation effect, without interference from other trivial bulk states. With such clean topological states, the Majorana states are supposed to be dominant in the superconducting gap. Distinct from previously studied platforms that support only a single Majorana mode, electron doped Fe(Te,Se) is a unique platform for exploring exotic topological superconducting phases and for investigating interactions between multiple Majorana modes.

\begin{addendum}

\item[Acknowledgements] We acknowledge Jianxin Li, Qianghua Wang, Jinsheng Wen, Kun Jiang, Xin Liu for useful discussions. This work was supported by the National Key R\&D Program of China (Grant No. 2024YFA1408400, 2024YFA1409100), the National Natural Science Foundation of China (Grant No. 12274209, 12374136), and the Fundamental Research Funds for the Central Universities (No. 14380230), the Natural Science Foundation of Jiangsu Province (No. BK20233001). Z.Z. acknowledges support from the Croucher Foundation through CIA23SC01 and the Hong Kong Research Grants Council through ECS 26308021.

\item[Correspondence] Correspondence and request for materials should be addressed to xxwu@itp.ac.cn, zhangpeng@nju.edu.cn.

\end{addendum}


\begin{thebibliography}{10}
\expandafter\ifx\csname url\endcsname\relax
  \def\url#1{\texttt{#1}}\fi
\expandafter\ifx\csname urlprefix\endcsname\relax\def\urlprefix{URL }\fi
\providecommand{\bibinfo}[2]{#2}
\providecommand{\eprint}[2][]{\url{#2}}

\bibitem{FuPRL2008}
\bibinfo{author}{Fu, L.} \& \bibinfo{author}{Kane, C.~L.}
\newblock \bibinfo{title}{{Superconducting Proximity Effect and Majorana
  Fermions at the Surface of a Topological Insulator}}.
\newblock \textit{\bibinfo{journal}{{Phys. Rev. Lett.}}}
  \textbf{\bibinfo{volume}{{100}}}, \bibinfo{pages}{{096407}}
  (\bibinfo{year}{{2008}}).

\bibitem{KouwenhovenScience2012}
\bibinfo{author}{Mourik, V.} \textit{et~al.}
\newblock \bibinfo{title}{{Signatures of Majorana Fermions in Hybrid
  Superconductor-Semiconductor Nanowire Devices}}.
\newblock \textit{\bibinfo{journal}{{Science}}}
  \textbf{\bibinfo{volume}{{336}}}, \bibinfo{pages}{{1003}}
  (\bibinfo{year}{{2012}}).

\bibitem{JiaPRL2016}
\bibinfo{author}{Sun, H.-H.} \textit{et~al.}
\newblock \bibinfo{title}{{Majorana Zero Mode Detected with Spin Selective
  Andreev Reflection in the Vortex of a Topological Superconductor}}.
\newblock \textit{\bibinfo{journal}{{Phys. Rev. Lett.}}}
  \textbf{\bibinfo{volume}{{116}}}, \bibinfo{pages}{{257003}}
  (\bibinfo{year}{{2016}}).

\bibitem{nadjperge2014observation}
\bibinfo{author}{Nadj-Perge, S.} \textit{et~al.}
\newblock \bibinfo{title}{Observation of majorana fermions in ferromagnetic
  atomic chains on a superconductor}.
\newblock \textit{\bibinfo{journal}{Science}} \textbf{\bibinfo{volume}{346}},
  \bibinfo{pages}{602--607} (\bibinfo{year}{2014}).

\bibitem{albrecht2016exponential}
\bibinfo{author}{Albrecht, S.~M.} \textit{et~al.}
\newblock \bibinfo{title}{Exponential protection of zero modes in majorana
  islands}.
\newblock \textit{\bibinfo{journal}{Nature}} \textbf{\bibinfo{volume}{531}},
  \bibinfo{pages}{206--209} (\bibinfo{year}{2016}).

\bibitem{ZhangScience2018}
\bibinfo{author}{Zhang, P.} \textit{et~al.}
\newblock \bibinfo{title}{Observation of topological superconductivity on the
  surface of an iron-based superconductor}.
\newblock \textit{\bibinfo{journal}{{Science}}} \textbf{\bibinfo{volume}{360}},
  \bibinfo{pages}{182} (\bibinfo{year}{2018}).

\bibitem{KanigelPRB2020}
\bibinfo{author}{Lohani, H.} \textit{et~al.}
\newblock \bibinfo{title}{{Band inversion and topology of the bulk electronic
  structure in FeSe$_{0.45}$Te$_{0.55}$}}.
\newblock \textit{\bibinfo{journal}{{Phys. Rev. B}}}
  \textbf{\bibinfo{volume}{{101}}}, \bibinfo{pages}{{245146}}
  (\bibinfo{year}{{2020}}).

\bibitem{JohnsonPNAS2021}
\bibinfo{author}{Zaki, N.}, \bibinfo{author}{Gu, G.}, \bibinfo{author}{Tsvelik,
  A.}, \bibinfo{author}{Wu, C.} \& \bibinfo{author}{Johnson, P.~D.}
\newblock \bibinfo{title}{{Time-reversal symmetry breaking in the
  Fe-chalcogenide superconductors}}.
\newblock \textit{\bibinfo{journal}{{Proc. Natl. Acad. Sci. U.S.A.}}}
  \textbf{\bibinfo{volume}{{118}}}, \bibinfo{pages}{{e2007241118}}
  (\bibinfo{year}{{2021}}).

\bibitem{LiNM2021}
\bibinfo{author}{Li, Y.} \textit{et~al.}
\newblock \bibinfo{title}{{Electronic properties of the bulk and surface states
  of Fe$_{1+y}$Se$_{1-x}$Te$_{x}$}}.
\newblock \textit{\bibinfo{journal}{{Nat. Mater.}}}
  \textbf{\bibinfo{volume}{{20}}}, \bibinfo{pages}{{1221}}
  (\bibinfo{year}{{2021}}).

\bibitem{ZXShenPRX2024}
\bibinfo{author}{Li, Y.-F.} \textit{et~al.}
\newblock \bibinfo{title}{{Orbital Ingredients and Persistent Dirac Surface
  State for the Topological Band Structure in FeTe$_{0.55}$Se$_{0.45}$}}.
\newblock \textit{\bibinfo{journal}{{Phys. Rev. X}}}
  \textbf{\bibinfo{volume}{{14}}}, \bibinfo{pages}{{021043}}
  (\bibinfo{year}{{2024}}).

\bibitem{DingScience2018}
\bibinfo{author}{Wang, D.} \textit{et~al.}
\newblock \bibinfo{title}{Evidence for majorana bound states in an iron-based
  superconductor}.
\newblock \textit{\bibinfo{journal}{Science}} \textbf{\bibinfo{volume}{362}},
  \bibinfo{pages}{333--335} (\bibinfo{year}{2018}).

\bibitem{HanaguriNM2019}
\bibinfo{author}{Machida, T.} \textit{et~al.}
\newblock \bibinfo{title}{{Zero-energy vortex bound state in the
  superconducting topological surface state of Fe(Se,Te)}}.
\newblock \textit{\bibinfo{journal}{Nat. Mater.}}
  \textbf{\bibinfo{volume}{18}}, \bibinfo{pages}{811} (\bibinfo{year}{2019}).

\bibitem{DingScience2020}
\bibinfo{author}{Zhu, S.} \textit{et~al.}
\newblock \bibinfo{title}{Nearly quantized conductance plateau of vortex zero
  mode in an iron-based superconductor}.
\newblock \textit{\bibinfo{journal}{Science}} \textbf{\bibinfo{volume}{367}},
  \bibinfo{pages}{189--192} (\bibinfo{year}{2020}).

\bibitem{KongNP2019}
\bibinfo{author}{Kong, L.} \textit{et~al.}
\newblock \bibinfo{title}{{Half-integer level shift of vortex bound states in
  an iron-based superconductor}}.
\newblock \textit{\bibinfo{journal}{{Nat. Phys.}}}
  \textbf{\bibinfo{volume}{{15}}}, \bibinfo{pages}{{1181}}
  (\bibinfo{year}{{2019}}).

\bibitem{GaoNature2022}
\bibinfo{author}{Li, M.} \textit{et~al.}
\newblock \bibinfo{title}{{Ordered and tunable Majorana-zero-mode lattice in
  naturally strained LiFeAs}}.
\newblock \textit{\bibinfo{journal}{Nature}} \textbf{\bibinfo{volume}{606}},
  \bibinfo{pages}{890--895} (\bibinfo{year}{2022}).

\bibitem{FengPRX2018}
\bibinfo{author}{Liu, Q.} \textit{et~al.}
\newblock \bibinfo{title}{{Robust and Clean Majorana Zero Mode in the Vortex
  Core of High-Temperature Superconductor (Li$_{0.84}$Fe$_{0.16}$)HOFeSe}}.
\newblock \textit{\bibinfo{journal}{{Phys. Rev. X}}}
  \textbf{\bibinfo{volume}{{8}}}, \bibinfo{pages}{{693}}
  (\bibinfo{year}{{2018}}).

\bibitem{DingNC2020}
\bibinfo{author}{Liu, W.} \textit{et~al.}
\newblock \bibinfo{title}{{A new Majorana platform in an Fe-As bilayer
  superconductor}}.
\newblock \textit{\bibinfo{journal}{Nat. Commun.}}
  \textbf{\bibinfo{volume}{11}}, \bibinfo{pages}{5688} (\bibinfo{year}{2020}).

\bibitem{DasSarmaPRL2019}
\bibinfo{author}{Zhang, R.-X.}, \bibinfo{author}{Cole, W.~S.} \&
  \bibinfo{author}{Das~Sarma, S.}
\newblock \bibinfo{title}{Helical hinge majorana modes in iron-based
  superconductors}.
\newblock \textit{\bibinfo{journal}{Physical Review Letters}}
  \textbf{\bibinfo{volume}{122}}, \bibinfo{pages}{187001}
  (\bibinfo{year}{2019}).

\bibitem{MadhavanScience2020}
\bibinfo{author}{Wang, Z.} \textit{et~al.}
\newblock \bibinfo{title}{Evidence for dispersing 1d majorana channels in an
  iron-based superconductor}.
\newblock \textit{\bibinfo{journal}{Science}} \textbf{\bibinfo{volume}{367}},
  \bibinfo{pages}{104--108} (\bibinfo{year}{2020}).

\bibitem{WangPRB2015}
\bibinfo{author}{Wang, Z.} \textit{et~al.}
\newblock \bibinfo{title}{{Topological nature of the FeSe$_{0.5}$Te$_{0.5}$
  superconductor}}.
\newblock \textit{\bibinfo{journal}{{Phys. Rev. B}}}
  \textbf{\bibinfo{volume}{{92}}}, \bibinfo{pages}{{115119}}
  (\bibinfo{year}{{2015}}).

\bibitem{WuPRB2016}
\bibinfo{author}{Wu, X.}, \bibinfo{author}{Qin, S.}, \bibinfo{author}{Liang,
  Y.}, \bibinfo{author}{Fan, H.} \& \bibinfo{author}{Hu, J.}
\newblock \bibinfo{title}{{Topological characters in Fe(Te$_{1-x}$Se$_{x}$)
  thin films}}.
\newblock \textit{\bibinfo{journal}{{Phys. Rev. B}}}
  \textbf{\bibinfo{volume}{{93}}}, \bibinfo{pages}{{115129}}
  (\bibinfo{year}{{2016}}).

\bibitem{ZhangNP2019}
\bibinfo{author}{Zhang, P.} \textit{et~al.}
\newblock \bibinfo{title}{Multiple topological states in iron-based
  superconductors}.
\newblock \textit{\bibinfo{journal}{Nat. Phys.}} \textbf{\bibinfo{volume}{15}},
  \bibinfo{pages}{41--47} (\bibinfo{year}{2019}).

\bibitem{JiaNature2024}
\bibinfo{author}{Liu, T.} \textit{et~al.}
\newblock \bibinfo{title}{Signatures of hybridization of multiple majorana zero
  modes in a vortex}.
\newblock \textit{\bibinfo{journal}{Nature}} \textbf{\bibinfo{volume}{633}},
  \bibinfo{pages}{71--76} (\bibinfo{year}{2024}).

\bibitem{HouPNAS}
\bibinfo{author}{Hou, Q.} \textit{et~al.}
\newblock \bibinfo{title}{{Bulk and surface Dirac states accompanied by two
  superconducting domes in FeSe-based superconductors}}.
\newblock \textit{\bibinfo{journal}{PNAS}} \textbf{\bibinfo{volume}{121}},
  \bibinfo{pages}{e2409756121} (\bibinfo{year}{2024}).

\bibitem{SunJPSJTeannealing}
\bibinfo{author}{Sun, Y.} \textit{et~al.}
\newblock \bibinfo{title}{{Bulk Superconductivity in
  Fe$_{1+y}$Te$_{1-x}$Se$_{x}$ Induced by Annealing in Se and S Vapor}}.
\newblock \textit{\bibinfo{journal}{J. Phys. Soc. Jpn.}}
  \textbf{\bibinfo{volume}{82}}, \bibinfo{pages}{093705}
  (\bibinfo{year}{2013}).

\bibitem{BezusyyPhysRevB.91.100502}
\bibinfo{author}{Bezusyy, V.~L.}, \bibinfo{author}{Gawryluk, D.~J.},
  \bibinfo{author}{Malinowski, A.} \& \bibinfo{author}{Cieplak, M.~Z.}
\newblock \bibinfo{title}{Transition-metal substitutions in iron
  chalcogenides}.
\newblock \textit{\bibinfo{journal}{Phys. Rev. B}}
  \textbf{\bibinfo{volume}{91}}, \bibinfo{pages}{100502}
  (\bibinfo{year}{2015}).

\bibitem{rosmus2020effect}
\bibinfo{author}{Rosmus, M.} \textit{et~al.}
\newblock \bibinfo{title}{{Effect of electron doping in FeTe$_{1-y}$Se$_{y}$
  realized by Co and Ni substitution}}.
\newblock \textit{\bibinfo{journal}{Superconductor Science and Technology}}
  \textbf{\bibinfo{volume}{32}}, \bibinfo{pages}{105009}
  (\bibinfo{year}{2019}).

\bibitem{supplement}
 \bibinfo{note}{{See Supplementary Informaiton for more details.}}

\bibitem{MiaoNC2015}
\bibinfo{author}{Miao, H.} \textit{et~al.}
\newblock \bibinfo{title}{{Observation of strong electron pairing on bands
  without Fermi surfaces in LiFe$_{1-x}$Co$_{x}$As}}.
\newblock \textit{\bibinfo{journal}{{Nat. Commun.}}}
  \textbf{\bibinfo{volume}{{6}}}, \bibinfo{pages}{{6056}}
  (\bibinfo{year}{{2015}}).

\bibitem{BorisenkoNP2016}
\bibinfo{author}{Borisenko, S.~V.} \textit{et~al.}
\newblock \bibinfo{title}{{Direct observation of spin--orbit coupling in
  iron-based superconductors}}.
\newblock \textit{\bibinfo{journal}{{Nat. Phys.}}}
  \textbf{\bibinfo{volume}{{12}}}, \bibinfo{pages}{{311}}
  (\bibinfo{year}{{2016}}).

\bibitem{WangPRB2012}
\bibinfo{author}{Wang, Z.} \textit{et~al.}
\newblock \bibinfo{title}{{Dirac semimetal and topological phase transitions in
  A$_3$Bi (A = Na, K, Rb)}}.
\newblock \textit{\bibinfo{journal}{{Phys. Rev. B}}}
  \textbf{\bibinfo{volume}{{85}}}, \bibinfo{pages}{{195320}}
  (\bibinfo{year}{{2012}}).

\bibitem{ChenScience2014}
\bibinfo{author}{Liu, Z.~K.} \textit{et~al.}
\newblock \bibinfo{title}{{Discovery of a Three-Dimensional Topological Dirac
  Semimetal, Na$_3$Bi}}.
\newblock \textit{\bibinfo{journal}{{Science}}}
  \textbf{\bibinfo{volume}{{343}}}, \bibinfo{pages}{{864}}
  (\bibinfo{year}{{2014}}).

\bibitem{JozwiakNC2016}
\bibinfo{author}{Jozwiak, C.} \textit{et~al.}
\newblock \bibinfo{title}{{Spin-polarized surface resonances accompanying
  topological surface state formation}}.
\newblock \textit{\bibinfo{journal}{{Nat. Commun.}}}
  \textbf{\bibinfo{volume}{{7}}}, \bibinfo{pages}{{106803}}
  (\bibinfo{year}{{2016}}).

\bibitem{PourretPRB2011}
\bibinfo{author}{Pourret, A.} \textit{et~al.}
\newblock \bibinfo{title}{{Strong correlation and low carrier density in
  Fe$_{1+y}$Te$_{0.6}$Se$_{0.4}$ as seen from its thermoelectric response}}.
\newblock \textit{\bibinfo{journal}{{Phys. Rev. B}}}
  \textbf{\bibinfo{volume}{{83}}}, \bibinfo{pages}{{020504}}
  (\bibinfo{year}{{2011}}).

\bibitem{KotliarNM2011}
\bibinfo{author}{Yin, Z.~P.}, \bibinfo{author}{Haule, K.} \&
  \bibinfo{author}{Kotliar, G.}
\newblock \bibinfo{title}{{Kinetic frustration and the nature of the magnetic
  and paramagnetic states in iron pnictides and iron chalcogenides}}.
\newblock \textit{\bibinfo{journal}{{Nature Mater}}}
  \textbf{\bibinfo{volume}{{10}}}, \bibinfo{pages}{{932}}
  (\bibinfo{year}{{2011}}).

\bibitem{ShenPRB2015}
\bibinfo{author}{Liu, Z.~K.} \textit{et~al.}
\newblock \bibinfo{title}{{Experimental observation of incoherent-coherent
  crossover and orbital-dependent band renormalization in iron chalcogenide
  superconductors}}.
\newblock \textit{\bibinfo{journal}{{Phys. Rev. B}}}
  \textbf{\bibinfo{volume}{{92}}}, \bibinfo{pages}{{235138}}
  (\bibinfo{year}{{2015}}).

\bibitem{KimPRL2024}
\bibinfo{author}{Kim, M.}, \bibinfo{author}{Choi, S.}, \bibinfo{author}{Brito,
  W.~H.} \& \bibinfo{author}{Kotliar, G.}
\newblock \bibinfo{title}{{Orbital-Selective Mott Transition Effects and
  Nontrivial Topology of Iron Chalcogenide}}.
\newblock \textit{\bibinfo{journal}{{Phys. Rev. Lett.}}}
  \textbf{\bibinfo{volume}{{132}}}, \bibinfo{pages}{{136504}}
  (\bibinfo{year}{{2024}}).

\bibitem{HuPRL2019}
\bibinfo{author}{Qin, S.} \textit{et~al.}
\newblock \bibinfo{title}{{Quasi-1D Topological Nodal Vortex Line Phase in
  Doped Superconducting 3D Dirac Semimetals}}.
\newblock \textit{\bibinfo{journal}{{Phys. Rev. Lett.}}}
  \textbf{\bibinfo{volume}{{123}}}, \bibinfo{pages}{{027003}}
  (\bibinfo{year}{{2019}}).

\bibitem{ColemanPRL2019}
\bibinfo{author}{K\"onig, E.~J.} \& \bibinfo{author}{Coleman, P.}
\newblock \bibinfo{title}{{Crystalline-Symmetry-Protected Helical Majorana
  Modes in the Iron Pnictides}}.
\newblock \textit{\bibinfo{journal}{{Phys. Rev. Lett.}}}
  \textbf{\bibinfo{volume}{{122}}}, \bibinfo{pages}{{207001}}
  (\bibinfo{year}{{2019}}).

\bibitem{ZhangarXiv2025}
\bibinfo{author}{Zhang, Z.} \textit{et~al.}
\newblock \bibinfo{title}{{Double Majorana Vortex Flat Bands in the Topological
  Dirac Superconductor}}  (\bibinfo{year}{{2025}}).
\newblock \eprint{{2501.05317v1}}.

\bibitem{SatoPRL2015}
\bibinfo{author}{Kobayashi, S.} \& \bibinfo{author}{Sato, M.}
\newblock \bibinfo{title}{{Topological Superconductivity in Dirac Semimetals}}.
\newblock \textit{\bibinfo{journal}{{Phys. Rev. Lett.}}}
  \textbf{\bibinfo{volume}{{115}}}, \bibinfo{pages}{{187001}}
  (\bibinfo{year}{{2015}}).

\end{thebibliography}
\end{document}